\documentclass[prb,twocolumn,showpacs,amsmath,amssymb]{revtex4}

\usepackage{graphicx}% Include figure files
\usepackage{dcolumn}% Align table columns on decimal point
\usepackage{bm}% bold math
\RequirePackage{color}
\include{revmacro}
\newcommand{\be}{\begin{equation}}
\newcommand{\ee}{\end{equation}}
\newcommand{\bea}{\begin{eqnarray}}
\newcommand{\eea}{\end{eqnarray}}

\begin{document}


%\title{Holmium films with a $J_1$-$J_2$ classical model}
\title{
%Helical orders and phase transitions in quasi-two-dimensional
%magnetic systems\\
%\textit{oppure}\\
% Competition among helical orders and surface effects in
% quasi-two-dimensional magnetic systems\\
% \textit{oppure}\\
Interplay among helical order, surface effects and range of 
interacting layers in ultrathin films.}

\author{F.~Cinti$^{(1,2,3)}$, A.~Rettori$^{(2,3)}$, and A.~Cuccoli$^{(2)}$}

\affiliation{$^{(1)}$ Department of Physics, University of Alberta, Edmonton, Alberta, Canada T6G 2J1}
\affiliation{$^{(2)}$CNISM and Department of Physics, University of Florence, 50019 Sesto Fiorentino (FI), Italy.}
\affiliation{$^{(3)}$CNR-INFM S$^3$ National Research Center, I-41100 Modena, Italy}

\date{\today}

\begin{abstract}
The properties of helical thin films 
have been thoroughly investigated by
classical Monte Carlo simulations. 
The employed model assumes classical planar spins 
in a body-centered tetragonal lattice, where the 
helical arrangement along the film growth direction 
has been modeled by nearest neighbor and 
next-nearest neighbor competing interactions, 
the minimal requirement to get helical order.
We obtain that,  while the in-plane transition temperatures 
remain essentially unchanged with respect to the bulk ones, 
the helical/fan arrangement is stabilized at more and more low
temperature when the film thickness, $n$, decreases; in the 
ordered phase, increasing the temperature, a softening of the 
helix pitch wave-vector is also observed.
Moreover, we show also that the simulation data around both transition 
temperatures lead us to exclude the presence of a first order 
transition for all analyzed sizes.
Finally, by comparing the results of the present work
with those obtained for other models previously adopted in 
literature, we can get a deeper insight about the entwined role 
played by the number (range) of interlayer interactions and surface 
effects in non-collinear thin films. 
\end{abstract}


\maketitle

% main text
\section{Introduction}\label{intro}

The study of low dimensional frustrated magnetic systems \cite{bookdiep}
still raises great interest, both in consequence of theoretical aspects,
related to their peculiar critical properties\cite{Kawamura}, and in view 
of possible technological applications \cite{app}.
Indeed, beside conventional ferromagnetic or antiferromagnetic phase 
transitions, in many new materials other nontrivial and unconventional 
forms of ordering have been observed\cite{catene,2dnem}.
A quantity of particular interest in this context is the 
spin chirality, an order parameter which turned out to be extremely 
relevant in, e.g., magnetoelectric materials \cite{mf1}, itinerant 
MnSi \cite{MnSi}, binary compounds as FeGe \cite{FeGe}, 
glass transition of spins \cite{kawa01}, and 
$XY$ helimagnets, as Holmium, Terbium or Dysprosium \cite{Jensen91}.
In the latter case, a new universality 
class was predicted because a $\mathbb{Z}_2\times SO(2)$ 
symmetry is spontaneously broken in the ordered phase \cite{Kawamura}:
In fact, when dealing with such systems, 
in addition to the $SO(2)$ symmetry of the spin degrees of freedom 
$\vec S_{i}$, one has to consider 
also the $\mathbb{Z}_2$ symmetry of the spin chirality 
$\kappa_{ij}\propto\left[\vec S_{i}\times\vec S_{j}\right]^z$.

For these rare-earth elements, the development
of new and sophisticated experimental methods \cite{Weschke08} has allowed
to obtain ultra-thin films where the non-collinear modulation  
is comparable with the film thickness. Under such conditions 
the lack of translational invariance due to the presence of surfaces
results decisive in order to observe a drastic change of the magnetic 
structures \cite{Jensen05}.
Recent experimental data on ultra-thin 
Holmium films \cite{Weschke04}  have been lately interpreted and discussed
 \cite{cinti1,cinti2} on the basis of detailed classical Monte
Carlo (MC) simulations of a spin Hamiltonian, which is believed to give
a realistic modeling of bulk Holmium. Such Hamiltonian, proposed by 
Bohr et al. \cite{Borh89}, allows for 
competitive middle-range interactions by including six different exchange 
constants along the $c$ crystallographic axis, and gives a helix pitch 
wave-vector $Q_z$ such that $Q_zc^\prime\simeq30^{\circ}$, where $c^\prime=c/2$ 
is the distance between nearest neighboring spin layers parallel to the $ab$ 
crystallographic planes, henceforth denoted also as $x-y$ planes, while 
$z$ will be taken parallel to $c$.
For $n\ >16$, $n$ being the number of spin layers in the
film, a correct bulk limit is reached, while for lower $n$ the film
properties are clearly affected by the strong competition among the helical
pitch and the surface effects, which involve the majority of the spin layers. In
the thickness range $n=9-16$, i.e. right for thickness values comparable with
the helical pitch, three different magnetic phases emerged, with the
high-temperature, disordered, paramagnetic phase and the low-temperature,
long-range ordered one separated by an intriguing, intermediate-temperature block
phase, where outer ordered layers coexist with some inner disordered ones, the
phase transition of the latter eventually displaying the signatures of a
Kosterlitz-Thouless one. Finally, for $n\leq7$ the film collapses once and 
for all to a quasi-collinear order.

The complex phase diagram unveiled by such MC simulations awaken  
however a further intriguing question: 
to  what extent the observed behavior may be considered a simple consequence 
of the competition between helical order and surface effects? I.e., is it just a matter
of having such a competition or does the range of interactions also play 
a relevant role? Indeed, when the range of the interactions is large enough
we have a greater number of planes which can be thought of as "surface planes",
i.e. for which the number of interacting neighbors are significantly reduced with
respect to the bulk layers; therefore, we expect that the larger the interaction range, 
the stronger should be the surface effects. But, at the same time, the same modulation of 
the magnetic order can be achieved with different number of interacting layers: notably,
nearest and next-nearest layers competitive interactions are enough to get a helical
structure with a whatever pitch wavevector. Such observation gives us a possible way 
to solve the conundrum previously emerged, as we have the possibility of varying the range
of interactions without modifying the helical pitch, thus decoupling the two
relevant length scales along the film growth direction, and making accessible a range 
of $n$ of the order of, or smaller than, the helical pitch, but still large enough
that a substantial number of layers can behave as ``bulk'' layers.  
Therefore, while in the previous papers we have studied the properties of ultrathin 
magnetic films of Ho assuming a model with six interlayer exchange 
interactions, here we investigate by MC simulations the properties of the same 
system by making use of the simplest model Hamiltonian able to describe the onset of a 
helical magnetic order in Holmium, i.e. we consider only two inter-layer coupling 
constants, as previously done in Ref. \onlinecite{Weschke08}. 

The paper is organized as follows: 
In Sec.~\ref{mmc} the model Hamiltonian will be defined, and 
the MC techniques, and all the thermodynamic quantities relevant
for this study, will be introduced. In Sec.~\ref{res} the results 
obtained for different thicknesses will be presented, both in
the matter of the critical properties of the model and of the 
magnetic ordered structures observed.
Finally, in Sec.~\ref{disc} we shall discuss such results,
drawing also some conclusions.

\section{Model Hamiltonian and Monte Carlo observables}\label{mmc}

\begin{figure}
\begin{center}
\includegraphics[width=0.4\textwidth]{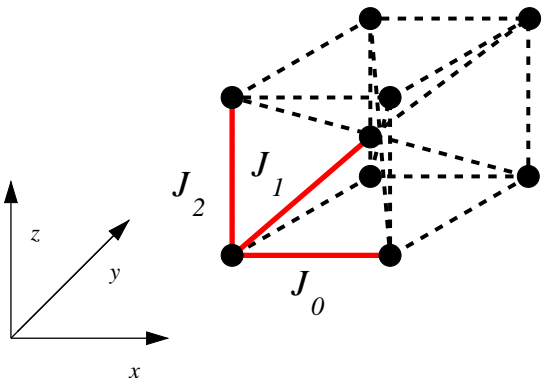}
\caption{\label{fig1} (colors online) \textbf{(a)}: body-centered
  tetragonal (BCT) lattice with $J_0$ in-plane coupling constant, and
  out-of-plane $J_1$, and $J_2$ competing interactions.}
\end{center}
\end{figure}

The model Hamiltonian we use in our simulations is the minimal one 
able to describe helimagnetic structures: 
% J_0 = 3.8280 / 4 = 0.9570 K 
% J_1 = 1.1600 / 4 = 0.2900 K
% J_2 = 0.3364 K    
\be
\label{ham}
{\cal H} = -\left[J_0\sum_{\langle ij \rangle} \vec{S}_{i}\cdot\vec{S}_{j}
+J_1\sum_{\langle ik \rangle} \vec{S}_{i}\cdot\vec{S}_{k}
+J_2\sum_{\langle il \rangle} \vec{S}_{i}\cdot\vec{S}_{l}\right] ~.
\ee
$\vec{S}_i$ are classical planar unit vectors representing the 
direction of the total angular momentum of the magnetic
ions, whose magnitude $\sqrt{j(j+1)}$ ($j=8$ for Holmium ions) 
is already encompassed within the definition of the interaction 
constants $J_{0,1,2}$. As sketched 
in Fig.~\ref{fig1}, the magnetic ions are
located on the sites of a body-centered tetragonal (BCT) lattice; 
the first sum appearing in the Hamiltonian describes the in-plane (\textit{xy}) 
nearest neighbor (NN) interaction, which is taken ferromagnetic (FM), 
with exchange strength $J_0>0$; the second sum 
represents the coupling, of exchange strength $J_1$, between spins 
belonging to nearest neighbor (NN) planes along the $z$-direction (which 
we will assume to coincide with the film growth direction); 
finally, the third sum takes into account the interaction, of exchange 
strength $J_2$, between spins lying on next-nearest
neighbor (NNN) planes along $z$. In order to have frustration, 
giving rise to non-collinear order along $z$ in the bulk, NN interaction 
$J_1$ can be taken both ferro- or antiferromagnetic, but NNN coupling $J_2$ 
has necessarily to be antiferromagnetic, and the condition $|J_2|>|J_1|/4$ must
be fulfilled. 
Such simplified Hamiltonian was already employed to simulate helical
ordering in bulk systems by Diep \cite{bookdiep,Diep89} and Loison 
\cite{Loison00b}.
In the bulk limit, the state of minimal energy of a system described
by Eq.(\ref{ham}) corresponds to a helical arrangement of spins. 
The ground state energy per 
spin is equal to $e_g(Q_z)=\left[-4J_0-2J_1\left(4\cos\left(Q_zc^\prime\right)+
\delta\cos\left(2Q_zc^\prime\right)\right)\right]$
where $c^\prime$ is the distance between NN layers, 
$\delta=\frac{J_2}{J_1}$, and $Q_zc^\prime=\arccos\left(-\frac{1}{\delta}\right)$
is the angle between spins lying on adjacent planes along the $z$-direction. 
The observed helical arrangement in bulk holmium corresponds to $Q_zc^\prime\simeq30.5^{\circ}$
\cite{Jensen91}: such value can be obtained from the formula above 
with the set of coupling constants $J_0$=67.2\,K, $J_1$=20.9\,K, 
and $J_2=-$24.2\,K, that we have employed in our simulations. The given values 
for the exchange constants are the same already used by  Weschke \textit{et al.}
in Ref.~\onlinecite{Weschke04} to interpret experimental data on Holmium 
films on the basis of a $J_1-J_2$ model, after a proper scaling by the numbers 
of NN and NNN on neighboring layers of a BCT lattice.

%In order to abtain a correct helimagnetic displacement of 
%the in-plane ($xy$) magnetic moments $J_1$ and $J_2$ have 
%to met the condition $|J_2|>J_1/4$, with $J_2$ necessarily AFM.
%The film structure is modelling
%taking along the $xy$-plane the FM interaction at NN, i.e.
%first sum of Hamiltonian~\eqref{ham}, $J_0$=0.957\,K,
%while the helix is developed along the $z$-direction exploit the
%competing between NN ($J_1$=0.29\,K) and NNN 
%($J_2=-$0.3364\,K) in the third sum.
%In Fig~\ref{fig1}b a representative picture for such helical
%structure is given. 
%This copling constant choice is able 
%to give an helix with step $\varphi=30.5^{\circ}$, as result by
%Mean-Field exitimation given in Refs.~\onlinecite{Weschke04}.

In the following we will denote with $n$ the film thickness, i.e.
the number of spin layers along the $z$ direction, 
and with $L$$\times$$L$ the number of spins in each layer 
(i.e., $L$ is the lattice size along both the $x$ and $y$ directions). 
In our simulations thickness values from 1 to 24 were considered, while
the range of lateral size $L$ was from 8 to 64. Periodic boundary 
conditions were applied along $x$ and $y$, while free boundaries
were obviously taken along the film growth direction $z$.

Thermal equilibrium was attained by the usual Metropolis 
algorithm\cite{met53}, supplemented by the over-relaxed 
technique\cite{over87} in order to speed-up the sampling 
of the spin configuration space: a typical ``Monte Carlo step'' 
was composed by four Metropolis and four-five 
over-relaxed moves per particle. Such judicious mix of moves 
is able both to get faster the thermal equilibrium
and to minimize the correlation ``time'' between successive
samples, i.e. the undesired effects due to lack of independence 
of different samples during the measurement stage.
For each temperature we have usually performed  
three independent simulations, each one containing at least
2$\times$10$^5$ measurements, taken after 
discarding up to 5$\times$10$^4$ Monte Carlo steps in order to
assure thermal equilibration.

In the proximity of the critical region 
the multiple histogram (MH) technique was also
employed \cite{Libbinder}, as it allows us to
estimate the physical observables of interest over 
a whole temperature range in a substantially continuous way
by interpolating results obtained from sets of simulations performed 
at some different temperatures. 

For all the quantities of interest, the average value and the error 
estimate were obtained by the bootstrap re-sampling 
method \cite{bootstrap} given that, as pointed out in 
Ref. \onlinecite{Efron79}, for a large enough number of measurements, 
this method turns out to be more accurate than the usual blocking technique.
In our implementation, we pick out randomly a sizable number of 
measurements (typically, between 1 and 1$\times$10$^{3}$ for the
single simulation, and between 1 and 5$\times$10$^{4}$ for the MH
technique), and iterate the re-sampling at least one hundred times.

The thermodynamic observables we have investigated 
include the FM order parameter for each plane $l$:
\be
\label{opsp}
m_l = \displaystyle{\sqrt{(m^x_l)^2 + (m^y_l)^2}}~~ ,
\ee   
which is related to the $SO(2)$ symmetry breaking. 
At the same time, it turns out to be significant also
the average order parameter of the film, defined as
\be
\label{opmp}
M = \frac{1}{n}\sum_{l=1}^{n} m_l\, .
\ee    
 
Turning to the helical order, which is the relevant quantity for 
the $\mathbb{Z}_2 \times SO(2)$ symmetry, we can explore it along two 
different directions. The first one is by the introduction of  
the chirality order parameter \cite{bookdiep,Kawamura} 
\be
\label{chirality}
\kappa = \frac{1}{4(n-1)L^2\sin Q_z}\sum_{\langle ij\rangle}
\left[S_i^xS_j^y-S_i^yS_j^x\right]\, ,
\ee
where the sum refers to spins belonging to NN layers $i$ and $j$,
respectively, while $Q_z$ is the bulk helical pitch vector along the 
$z$ direction. 
The second possibility is that of looking at the integral of the
structure factor:
\be
\label{mq}
M_{HM}=\frac{1}{K}\int_0^\pi dq_z S(\vec{q})
\ee
where $S(\vec{q})$, with $\vec{q}=(0,0,q_z)$, 
is the structure factor \cite{lubbook} (i.e. the Fourier transform of the 
spin correlation function) along the z-direction of the film,
while the normalization factor $K$ is the structure factor integral at $T=0$.
Although the use of the last observable can be seen as a suitable
and elegant way to overcome the intrinsic difficulties met in defining
a correct helical order parameter, free of any undue external 
bias (as the wave-vector $Q_z$ entering the definition of $\kappa$ 
in Eq.~(\ref{chirality})), we remind that such quantity has generally 
to be managed with particular care, as discussed in details 
in Refs.~\cite{cinti1,cinti2}, where it was shown that the presence 
of block structures prevents us to unambiguously relate the evolution 
of $S(\vec{q})$ with the onset of helical order.
However, 
%taking into account that the Hamiltonian~\eqref{ham}
%has competing interactions which extend up to NNN only,
for the specific case of the model under investigation such integrated 
quantity can still be considered a fairly 
significant order parameter, as no block structures emerge from
the simulations (see below). 

%This is probably due  
%as soon as $n$ is moderately large,  .

In order to get a clear picture of the critical region
and to give an accurate estimate of the critical temperature,
we look also at the following quantities 
\bea
c_v &=&nL^2\beta^2\left(\langle e^2\rangle -\langle e
  \rangle^2\right)\, , \label{cv}\\
\chi_o &=& nL^2\beta \left(\langle o^2 \rangle -\langle o
  \rangle^2\right)\, ,\label{chi}\\
\partial_\beta o  &=& 
nL^2\left(\langle oe\rangle-\langle o\rangle\langle e \rangle\right)\, , \label{derk}\\
u_{4}(o) &=& 1 - \frac{\langle o^4 \rangle}{3\langle o^2 \rangle^2}\, ,\label{u4g}
\eea
where $\beta=1/k_BT$, and $o$ is one of the relevant observables, i.e. 
$m_l,M,\kappa,M_{HM}$.
In this paper, we shall mainly locate the critical temperature by looking at the 
intersection of the graphs of the Binder cumulant \cite{binderu4}, 
Eq.~\eqref{u4g}, as a function of $T$ obtained at different $L$. 
For clarity reasons, we introduce also the following symbols: by $T_N(n)$ 
we will denote the helical/fan phase transition temperature for thickness $n$,
$T_C(n)$ will instead indicate the ordering temperature of the sample 
as deduced by looking at the behaviour of the 
average order parameter~\eqref{opmp}, while $T_C^l(n)$ will be 
the $l$-th plane transition temperature related to the order parameter defined in   Eq.~\eqref{opsp}.

\section{Results}\label{res}

\begin{figure}
\begin{center}
\includegraphics[width=0.4\textwidth]{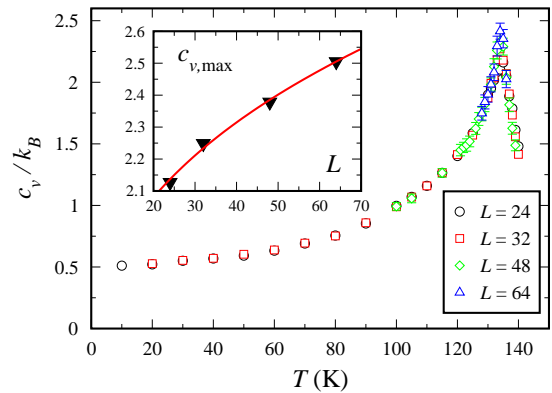}
\caption{\label{cv16} (color online) Specific heat $c_v$ per spin
  vs. temperature for thickness $n=16$ (for lateral dimension, see the
  legend inside the figure). Inset: Maximum of $c_v$ vs. $L$
  obtained through MH technique. The continuum red line is a power law
  fit. %Fit value: $C_0$=1.10487,  $C_1$=0.384008,  and
       %$\alpha/\nu$=0.311085.
}
\end{center}
\end{figure}

The results obtained by MC simulations of the model introduced 
in Sec.~\ref{mmc} will be presented starting from $n=16$, i.e. the highest 
investigated film thickness which still displays a bulk-like behaviour. 
In  Fig.~\ref{cv16} the specific heat for samples with $n=16$
and lateral dimension $L=24,\,32,\,48,\,64$ is shown. 
The location of the specific heat maximum shows 
a quite definite evolution toward the bulk 
transition temperature, $T_N^{Ho}\simeq132$\,K \cite{Jensen91} 
(it is worthwhile to note that for this $XY$ model the mean 
field theory predicts a critical temperature 
$T_{N,MF}^{Ho}\simeq198$\,K).
 
The intensity of the maximum of 
$c_v$ has been analyzed by the MH technique for the same lateral dimensions
(see inset of Fig.~\ref{cv16}): it clearly appears as it increases with $L$ 
in a smooth way.

\begin{figure}
\begin{center}
\includegraphics[width=0.4\textwidth]{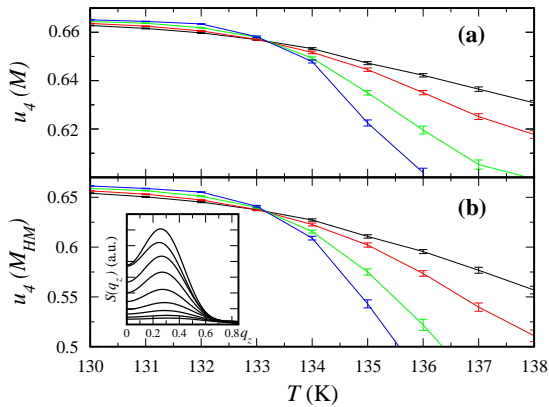}
\caption{\label{u416} (color online) Binder cumulants at thickness $n=16$, 
  colors as in Fig.~\ref{cv16}. \textbf{(a)}: Binder cumulant
  for the order parameter defined in Eq.~\eqref{opmp}. \textbf{(b)}:
  Binder cumulant extracted from the integral of the structure factor (see
  Sec.~\ref{mmc}). Inset: structure factor for $L=64$ between $T=131$~K 
  (upper curve) and $T=140$~K (lower), with $1$ K temperature step.}
\end{center}
\end{figure}
The Binder cumulant for the average order parameter defined in Eq.~\eqref{opmp} 
was obtained close to the $c_v$ peak and is reported in Fig.~\ref{u416}a;
its analysis leads to an estimate of the critical temperature of the sample
(given by the location of the common crossing point of the different curves reported 
in the figure) of $T_C(16)=133.2(5)$ 
This value can be considered in a rather good agreement with the experimental ordering 
temperature of Holmium $T_N^{Ho}$, the relative difference being about 
1\%. Even such a mismatch between $T_N^{Ho}$ and $T_C(16)$ could be
completely eliminated by slightly adjusting the in-plane coupling
constant $J_0$, but, as discussed in Sec.~\ref{mmc}, we shall
preserve the value reported in Refs.~\onlinecite{Weschke04}, and~\onlinecite{Jensen05} 
in order to allow for a correct comparison with the results reported in those papers. 

The development of the helical arrangement of magnetization along the film growth 
direction was investigated by looking at the integral of the structure factor 
$S(\vec{q})$ along the $z$-direction, i.e. by taking $\vec{q}=(0,0,q_z)$, 
and making again use of the cumulant analysis in order to locate the helical 
transition temperature at $T_N(16)=133.1(3)$\,K (see Fig.~\ref{u416}b). 
The crossing points of the Binder's cumulants of the helical order parameter 
immediately appear to be located, within the error bars, at the same 
temperature of those for the average magnetization previously discussed.
In addition, it is worthwhile to observe that the peak evolution 
of $S(0,0,q_z)$, in particular close to $T_N(16)$ (inset of Fig.~\ref{u416}b), 
displays the typical behaviour expected for an helical structure.
We can thus conclude that for $n=16$, as it is commonly observed in bulk samples,
the establishment of the in-plane order coincides with onset of the perpendicular 
helical arrangement at $T_N(16)$. However, due to helix distortion in the surface 
regions, the maximum of $S(0,0,q_z)$ stabilizes at values of $q_z$ sensibly
smaller (e.g. $Q_z(T_N(16))\thickapprox 16^\circ$, and
$Q_z(T=10K)\thickapprox 28^\circ$) with respect to the bulk
one ($Q_z^{Ho}=30.5^\circ$). 
 
The MC simulations outcomes for $n=16$ we just presented  
appear quite different with respect to those obtained
at the same thickness for the model with six coupling constants
along the $z$  direction \cite{cinti1,cinti2}. 
Indeed, for the $J_1$-$J_2$ model here investigated, we observe 
that all layers order at the same temperature, and we do not find any hint 
of the block-phase, with inner disordered planes intercalated to
antiparallel \textit{quasi}-FM four-layer blocks, previously observed;
sample MC runs we made using the same $hcp$ lattice employed in 
Refs.~\onlinecite{cinti1, cinti2} shows that the presence or absence of the
block phase is not related to the lattice geometry, but it is a consequence
of the interaction range only.
%So for $n=16$ the surface effects do not are not dominate by the
%magenetic critical properties with respect to the precedent works. 

\begin{figure}
\begin{center}
\includegraphics[width=0.5\textwidth]{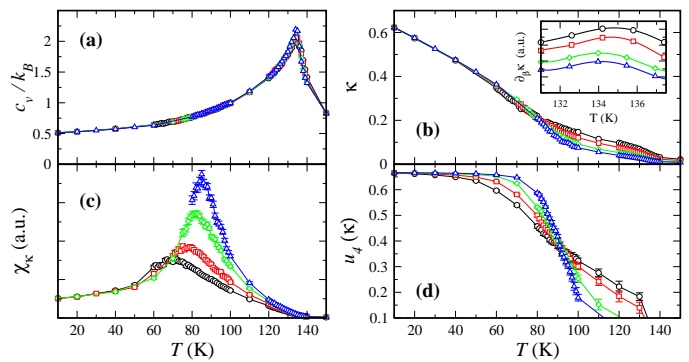}
\caption{\label{sin08} (color online) Thermodynamic quantities obtained
  for thickness $n=8$ in the temperature range 0-150\,K. Colors and 
  symbols as in Fig.~\ref{cv16}. 
  \textbf{(a)}: specific heat; \textbf{(b)}:
  chirality order parameter. 
  \textbf{(c)}: susceptibility
  $\chi_\kappa$. \textbf{(d)}: Binder cumulant for $\kappa$.}
\end{center}
 \end{figure}
\begin{figure}
\begin{center}
\includegraphics[width=0.4\textwidth]{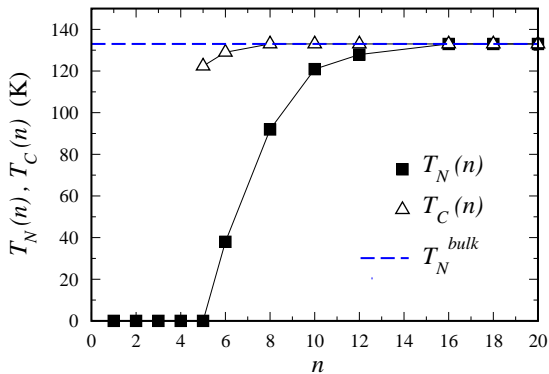}
\caption{\label{tvsn} Transition temperatures $T_N(n)$ and  
$T_C(n)$ vs. film thickness $n$.}
\end{center}
\end{figure}

We now move to describe and discuss MC simulation data for thinner samples.
A graphical synthesis of the results obtained for $n=8$ in reported in 
Fig.~\ref{sin08}a-d. The specific heat $c_v$, shown in Figs.~\ref{sin08}a,
reveals very small finite-size effects,
which, however, cannot be unambiguously detected for
the largest lattice size ($L=64$), as they fall comfortably
within the error range. Surprisingly, the specific heat maximum
is located close to the bulk transition temperature as found for
$n=16$, and the same is true for the crossing point of the Binder cumulant 
of the average magnetization $M$ (not reported in figure), which is located
at  $T_C(8)=133.3(3)$\,K. These data give a first rough indication that 
also for $n=8$ all the planes of the sample are still ordering almost at the same 
temperature; such property has been observed for all the investigated 
thicknesses $n$ below $16$, so that $T_C(n)$ results quite $n$-independent 
(see also Fig.~\ref{tvsn}) .

Although the layer subtraction does not seem to modify $T_C(n)$,
the onset of helical arrangement is observed to shift at lower temperatures as
$n$ decreases.
The chirality $\kappa$ defined in Eq.~(\ref{chirality}) is reported in Fig~\ref{sin08}b 
for $n=8$. 
As the temperature decreases, around $T\sim 80$\,K we can identify a finite-size 
behaviour of $\kappa$ which, at variance with the previous one, can be easily 
recognized as typical of an effective phase transition.
Such conclusion is confirmed by the analysis of the chiral susceptibility $\chi_\kappa$ 
(Fig.~\ref{sin08}c), which for the largest $L$ has a maximum
at $T=85$\,K. 
Assuming that the order parameter~\eqref{chirality} is the relevant one to 
single out the onset of the fan arrangement, we can get a more accurate 
estimate of $T_N(8)$ by looking at the Binder cumulant $u_{4}(\kappa)$, 
reported in Fig.~\ref{sin08}d. By making use of the MH technique, 
we locate the crossing point at $T_N(8)=92(2)$\,K.
Finally, it is worthwhile to observe as the specific heat 
does not show any anomaly at $T_N(8)$, being the entropy
substantially removed at $T_C(8)$.

The scenario just outlined for $n=8$ results to be correct in the
thickness range $6\le n \lesssim 15$, where a clear separation 
between $T_N(n)$ and $T_C(n)$ can be easily figured out.
In such temperature window, the strong surface effects produce 
a \textit{quasi}-FM set-up of the magnetic film structure
along the $z$-direction.
While leaving to the next Section a more detailed discussion of this regime,
we report in Fig.~\ref{tvsn} a plot of $T_N(n)$ and $T_C(n)$ vs. $n$ for all
the simulated thicknesses. The separation between the two critical temperatures
is maximum for $n=6$, where $T_N(6)=38(4)$, that is $T_N(6)\sim\frac{1}{3}T_C(6)$.
For films with less than six layers no fan order is observed, i.e. 
for $n=5$ and below the chirality does not display any typical feature 
of fan ordering at any temperature below $T_C(n)$. 
\begin{figure}
\begin{center}
\includegraphics[width=0.4\textwidth]{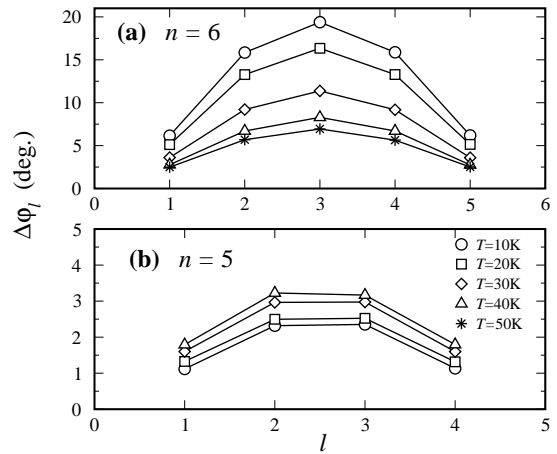}
\caption{\label{dphi}  
  Rotation angle $\Delta\varphi_l$ 
  between magnetic moments on NN layers $(l+1,l)$ at some low temperatures,
  for thickness $n=5$ and $n=6$, and lateral dimension $L=64$.}
\end{center}
\end{figure}
As a representative quantity we finally
look at the rotation angle of the magnetization between nearest planes:
\be
\label{dphi0}
\Delta\varphi_l=\varphi_{l+1}-\varphi_{l}=\arccos\left[M^x_lM^x_{l+1}+M^y_lM^y_{l+1}\right]
\ee
where $(M^x_l,M^y_l)$ is the magnetic vector profile for each plane $l$.
$\Delta\varphi_l$ is displayed in Fig.~\ref{dphi}a and
Fig.~\ref{dphi}b, for $n=6$ and $n=5$, respectively. 
In Fig.~\ref{dphi}a, a quite clear fan stabilization is observed
when the temperature decreases, while in Fig.~\ref{dphi}b, i.e. for 
$n=5$, $\Delta\varphi_l$ keeps an almost temperature independent 
very small value; what's more, $\Delta\varphi_l$ seems to loose
any temperature dependence as $T=0$ is approached.
We attribute the absence of fan arrangement for $n\le 5$ as
simply due to the lack of ``bulk planes'' inside the film, so that we 
are left with only a 2d trend at $T_C(n)$, i.e. at the temperature where 
the order parameters defined in Eqs.~\eqref{opsp} and \eqref{opmp} show a critical
behaviour.

%\textbf{!!! magari qui devo spiegare meglio: essendo $M$ legato ai
%piani FM osservero' qualcosa che sara' sempre piu' 2d. Infatti per 
%gli spessori pi\`u bassi ($n\le 8$) ho analizzato il
%$\ln\chi_M$ vs. $\ln L$, ottendo un $\gamma/\nu$ sempre vicino a
%1.75 (nel limite dell'errore) !!! {\color{red} Se \`e cos\'\i~ direi che sarebbe
%bene aggiungerlo}}.
%\textbf{!!! devo fare un fit della Figura~\ref{tvsn} !!! {\color{red} Cio\`e ????}}

\section{Discussion and Conclusion}\label{disc}

A possible framework to analyze the results presented in the previous 
Section is suggested by Fig.~\ref{tvsn}, where we can easily distinguish 
three significant regions: \textit{i}) high thickness, 
$n\geqslant 16$, where the films substantially display a bulk behaviour, 
with the single planes ordering temperature coinciding with the helical
 phase transition one;
\textit{ii}) intermediate thickness, $6\le n \lesssim 15$, where the 
temperature corresponding to the onset of in-plane order, $T_C(n)$, is 
still $\simeq T_N^{Ho}$, but where the helical/fan arrangement stabilizes 
only below a finite temperature $T_N(n)<T_C(n)$; \textit{iii}) low
thickness,$1\le n \le 5$,  where $T_C(n)\lesssim T_N^{Ho}$ but no fan 
phase is present at any temperature.

%%%%%%%%%%%%%%%%%%%%%%%%%%%%%%%%%%%%%%%%
%%%%%%%%%%%%%%%%%%%%%%%%%%%%%%%%%%%%%%%%
%%%%%%%%%%%%%%%%%%%%%%%%%%%%%%%%%%%%%%%%
The observed behaviour in region \textit{iii}) can be reasonably attributed 
to the decreasing relevance of the contribution to the total energy of the 
system coming from the competitive interactions among NNN planes as the film
thickness decreases; moreover, the thinness of the film leads to an effective 
2d-like trend. Region \textit{ii}) 
looks however more intriguing, and requires a more accurate discussion, which can 
benefit from a careful comparison of the behaviour of a given quantity in regions 
\textit{i}) and \textit{ii}).
\begin{figure}
\begin{center}
\includegraphics[width=0.4\textwidth]{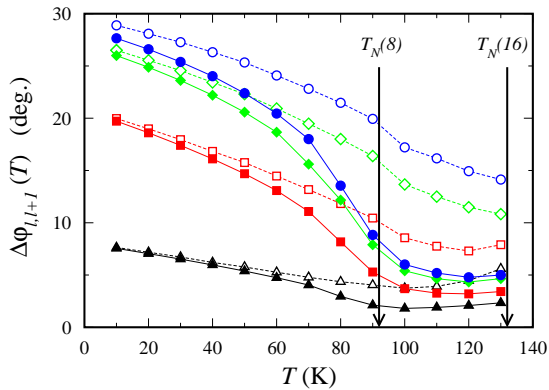}
\caption{\label{dphiT}(color online) $\Delta\varphi_l(T)$ vs. temperature 
     for the surface planes, $l=1$ (triangles), $l=2$ (squares), 
     $l=3$ (diamonds), $l=4$ (circles). Straight lines and full symbols: 
     $n=8$. Dashed lines and open symbols: $n=16$.}
\end{center}
\end{figure}

For this purpose, we look at the temperature dependence of the rotation angle 
of the magnetization between NN planes. In Fig.~\ref{dphiT},  $\Delta\varphi_l(T)$ 
for $n=8$ and $n=16$ (continuous and dashed lines, respectively), is plotted 
for the outermost planes, $l=1\dots4$. For both thicknesses, a monotonic trend 
is observed for all $l$, but at variance with what happens for the highest
thickness, for $n=8$ we see, starting from a temperature $T\lesssim T_N(8)$, 
an abrupt drop of $\Delta\varphi_{3}$ and $\Delta\varphi_{4}$, which rapidly 
reach an almost constant value, only slightly 
larger than $\Delta\varphi_{1}$. In the temperature range $T_N(8)\lesssim T<T_C(8)$ we
thus substantially observe the same small magnetic phase shifts between all NN layers, 
testifying an energetically stable \textit{quasi}-FM configuration giving no contribution
to the helical order parameters. 
\begin{figure}
\begin{center}
\includegraphics[width=0.4\textwidth]{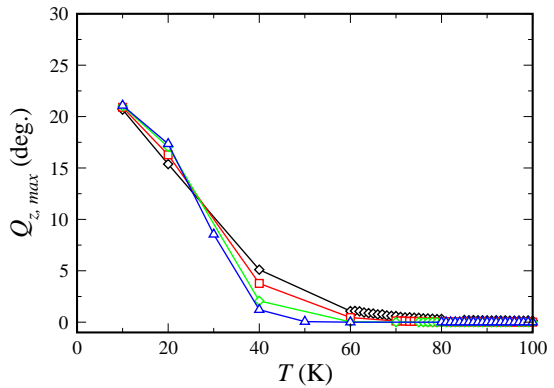}
\caption{\label{figqmax} (color online) $Q_z$, position of the maximum of 
$S(\vec{q})$, vs. temperature for thickness
  $n=8$. Inset: magnetic vector $(m^x_l,m^y_l)$ profile for some
  temperatures for $L=64$. Colors and symbols as in Fig.~\ref{cv16}.}
\end{center}
\end{figure}
%For the case (\textit{ii}),  we have formerly suggested that in the temperature range
%$T_N(n)<T<T_C(n)$ the surface effects produce a magnetic configuration 
%where the in-plane magnetic moments find a stable \textit{quasi}-FM set-up.
The latter point can be made clearer by looking at the the peak position $Q_{z,max}$
of the structure factor $S(0,0,q_z)$.
In Fig.~\ref{figqmax} the average of $Q_{z,max}$ $vs$ $T$ is reported, again for $n=8$
and for different lateral dimensions $L$ \footnote{Such observable 
has been obtained from instantaneous evaluation
of the structure factor during the stochastic process, 
and subsequently statistically analyzed as all the other macroscopic quantities.}.
As expected from the previous argument, we see that $Q_{z,max}=0$ for $T_N(8)<T<T_C(8)$, 
while it begins to shift to higher values as soon as the temperature decreases below $T_N(8)$, 
making apparent a progressive fan stabilization with $Q_{z,max}\ne 0$ and reaching a value 
of about $21^\circ$ for $T=10$\,K. 
%{\bf Togliere l'inset di Fig. 8}.
% Some examples of the magnetic 
% vector profile $(M^x_l,M^y_l)$ along 
% the film growth direction are shown in the inset of 
% Fig.~\ref{figqmax} for some temperatures below 
% $T_C(8)$.
%%%%%%%%%%%%%%%%%%%%%%%%%%%%%%%%%%%%%%
%%%%%%%%%%%%%%%%%%%%%%%%%%%%%%%%%%%%%%
%%%%%%%%%%%%%%%%%%%%%%%%%%%%%%%%%%%%%%

\begin{figure}
\begin{center}
\includegraphics[width=0.4\textwidth]{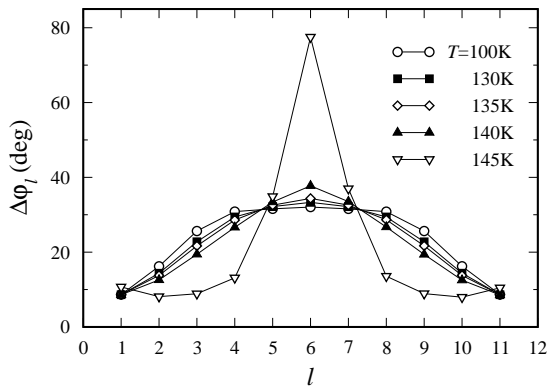}
\caption{\label{dphi12} $\Delta\varphi_l$ for a BCT lattice and 
$n = 12$, when the six coupling constants set employed in 
Ref.~\onlinecite{cinti1,cinti2} (see text) is used.
The temperature range has been chosen around $T_C(n)$
(error bars lye within point size).}
\end{center}
\end{figure}
In a previous study, where the magnetic properties of Ho thin films 
were investigated by MC simulations of a Heisenberg model with 
easy-plane single-ion anisotropy and six out-of-plane coupling
constants (as obtained by experimental neutron scattering measurements \cite{Borh89})
on a HCP lattice \cite{cinti1,cinti2}, it was found that for thicknesses comparable with the helical pitch the phase diagram landscape is quite different 
from what we find here. Indeed, for $n=9-16$, 
three different magnetic phases could be singled out, with the high-temperature,
paramagnetic phase separated from the low-temperature, long-range ordered one, by an 
intermediate-temperature block phase where outer ordered 4-layers blocks coexist with some
inner disordered ones. Moreover, it was observed that the phase transition of such
inner layers turns out to have the signatures of a Kosterlitz-Thouless one.

The absence of the block phase in 
the $J_1-J_2$ model 
here investigated has to be attributed to the different range of interactions, 
rather than to the different lattice structure. We came to this conclusion
by doing some simulations using the same set of interaction 
constants employed in Refs.~\onlinecite{cinti1,cinti2}, but using a BCT lattice:
the results we obtained for $\Delta\varphi_l$ with $n=12$ are reported 
in Fig.~\ref{dphi12}. The latter is absolutely similar to Fig.7 of 
Ref.~\onlinecite{cinti2} and clearly 
displays the footmarks of the block phase (see down-triangle), with two external
blocks of ordered layers ( $l=$1\dots5  and 8\ldots12 ), where  $\Delta\varphi_l$ 
is roughly $10^\circ$, separated by a block of disordered layers, and with almost
opposite magnetization. 
We can thus confidently assert that, regardless of the underlying lattice structure,
by decreasing the number of the out-of-plane interactions,
for thicknesses close to the helical bulk pitch, the block phase is 
replaced by a \textit{quasi}-FM configuration in the intermediate temperature 
range $T_N(n)<T<T_C(n)$ .

\begin{figure}
\begin{center}
\includegraphics[width=0.4\textwidth]{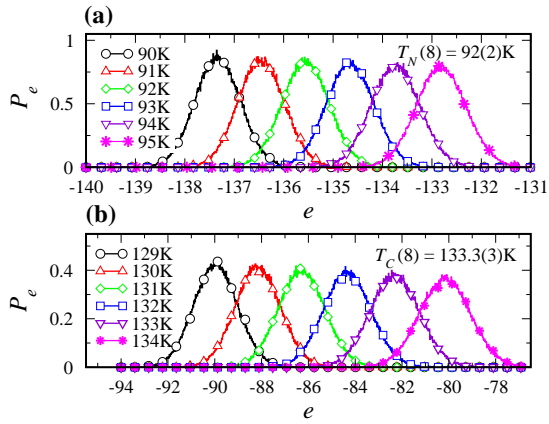}
\caption{\label{histo8} 
(colors online) Equilibrium probability distribution of the energy for the thickness $n=8$
for some temperatures around $T_N(8)$, \textbf{(a)}, and $T_C(8)$, 
\textbf{(b)}, respectively.}
\end{center}
\end{figure}
As a final issue we address the problem of the order of the transitions
observed at $T_N(n)$ and $T_C(n)$, respectively. 
In particular, we focus our attention to the thickness ranges where the chiral
order parameter is relevant, i.e. regions \textit{i}) and \textit{ii}) as defined 
at the beginning of this Section.
In Fig.~\ref{histo8} the equilibrium probability distribution of the energy for 
temperatures around $T_N(8)$ (Fig.~\ref{histo8}a) and $T_C(8)$ (Fig.~\ref{histo8}b) 
is plotted: for both temperatures, no double peak structure is observed, so that we
have no direct indication for a first order transition even if, according to precedent
studies of Loison and Diep \cite{Loison00b,Diep89}, the presence of a first-order transition at $T_N(n)$,  
cannot be completely excluded, as it could reveal itself only 
when the lateral dimension $L$ are much larger than the largest correlation
length. 
The same conclusion about the order of transition is reached
for any other investigated film thickness, as the energy probability distribution shape 
does not qualitatively change. This findings agree with the results we got in previous 
MC simulations discussed in Ref.~\onlinecite{cinti2}, so that we may conclude that 
the order of the observed transitions is not affected by the range of interactions.

%\section{Conclusion}\label{conc}

% \appendix

% \section{}
% \label{}

% Bibliographic references with the natbib package:
% Parenthetical: \citep{Bai92} produces (Bailyn 1992).
% Textual: \citet{Bai95} produces Bailyn et al. (1995).
% An affix and part of a reference:
%   \citep[e.g.][Ch. 2]{Bar76}
%   produces (e.g. Barnes et al. 1976, Ch. 2).

%{\bf Le citazioni devono essere rimesse in ordine}

\end{document}